\documentclass[12pt,aps,prb,showpacs,preprint, english]{revtex4}
\usepackage{graphicx}
\usepackage{dcolumn}
\usepackage{amsmath}
\usepackage{amssymb}
\usepackage{times}
\usepackage{babel}
\usepackage{epstopdf}
\usepackage{color}
\usepackage{epstopdf}

\def\tV{{\tilde V}}

\def\bfr{{\bf r}}

\def\bfs{{\bf s}}

\def\bfq{{\bf q}}

\def\beq{\begin{equation}}
\def\eeq{\end{equation}}

\begin{document}
\title{Gradient corrections to the kinetic energy density functional of a two-dimensional Fermi gas at finite temperature}
\author{B. P. van Zyl$^{(1)}$, K. Berkane$^{(2)}$, K Bencheikh$^{(2)}$, A. Farrell$^{(1)}$}
\affiliation{$^{1}$Department of Physics, St. Francis Xavier University, 
Antigonish, NS, Canada B2G 2W5} 
\affiliation{$^{2}$D\'{e}partement de Physique. Laboratoire de physique quantique et syst%
\`{e}mes dynamiques. Universit\'{e} de S\'{e}tif, Setif 19000, Algeria.}
\date{\today}

\pacs{71.15.Mb,03.65.Sq,05.30.Fk,31.15.Bs}

\begin{abstract}
We examine the leading order semiclassical gradient corrections to the non-interacting kinetic energy density functional of a two dimensional Fermi gas by applying the extended Thomas-Fermi theory
at finite temperature.  We find a non-zero von Weizs\"acker-like gradient correction, which
in the high-temperature limit, goes over to the functional form
$(\hbar^2/24m) (\nabla\rho)^2/\rho$.  Our  work provides a theoretical justification for the inclusion of  gradient corrections in applications of density-functional
theory to inhomogeneous two-dimensional Fermi systems at any {\em finite} temperature.
\end{abstract}
\maketitle
\section{Introduction}
In 1966, Fower {\it et al.}~\cite{fowler} performed transport measurements on a Si-metal-oxide-semiconductor structure in which a degenerate gas of electrons was electrostatically
induced.   Their work demonstrated for the first
time that the density of states in the $n$-type electron inversion layer had the expected behaviour for a two-dimensional electron gas (2DEG).  Since Fowler's seminal work,
the exploitation of the electronic properties of III-V semiconductors has led to the realization of high quality, high mobility 2DEG's at the interface of epitaxially grown III-V 
structures, such as GaAs/AlGaAs heterostructures.~\cite{solid_state}   Through electrostatic, and/or etching techniques, the 2DEG found in the III-V semiconductor interface, can be manipulated to 
create experimental realizations of low dimensional
electron systems such as quantum wires, quantum dots and quantum anti-dots.

By far, the workhorse for a theoretical understanding of the bulk electronic properties of such low-dimensional electronic systems is the zero-temperature ($T=0$) density-functional theory (DFT) of Hohenberg, Kohn  and Sham (HKS).~\cite{hohenberg,ks}  The key element in the HKS approach is the definition of the kinetic energy (KE), corresponding to a system of $N$ noninteracting fermions 
moving in some
effective one-body potential.  The HKS scheme treats the KE {\em exactly} at the independent particle level, and so the development of explicit, orbital-free
functionals for ${\cal T}[\rho]$, the KE density functional, is an important objective.  Ideally the appropriate functional should
yield both the correct energy {\em and} the correct density profile.

To this end, the simplest approach for the construction of the KE density functional, ${\cal T}[\rho]$, is the local-density approximation (LDA), sometimes referred to as the Thomas-Fermi (TF)
approximation.~\cite{thomas,fermi}  In this approximation, the {\em known} form for KE density functional of the {\em uniform} electron gas, is also used locally for the KE density functional of the {\em inhomogeneous}
system.
One would then expect the LDA to be applicable only in cases
where the density in the system is a slowly varying function of position.  In fact, this is not the case, and even in highly inhomogeneous systems, the LDA is found to work
reasonably well.~\cite{dft2,dft1}

Although the LDA for the KE density leads to reasonable results for the energy, the calculated density profile in a self-consistent DFT scheme does not exhibit the desired quantum mechanical tunnelling into the classically forbidden region.  To overcome this issue, the so-called von Weizs\"acker (vW) gradient correction,~\cite{vW} $\propto (\nabla \rho)^2/\rho$, is added to
the KE functional.  In 3D, the vW gradient correction can be rigorously justified within the extended TF (ETF) theory,
originally developed in the context of nuclear physics.~\cite{GGA,brack_bhaduri}  The inclusion of the vW term leads to smooth and continuous densities, while improving the quality of the KE functional by taking into account the inhomogeneity of the system.

An application of the ETF theory to 2D systems, however, leads to the conclusion that there are {\em no gradient
corrections} to the 2D KE density functional.~\cite{march,salasnich,koivisto}  This, of course, makes no physical sense since the LDA cannot be  variationally exact for an inhomogeneous system.
Thus in DFT applications to systems derived from the inhomogeneous 2DEG discussed above, a phenomenological approach must be taken in which a vW-like
gradient correction is put in ``by hand''.~\cite{vz1}
Although the vW-like correction term is entirely {\it ad hoc} for a 2D system, its use has been justified by (i) the KE reduces
to the TF limit for slowly varying densities, and (ii) it allows one to represent strongly inhomogeneous densities in a quantum mechanically reasonable way. 

To date, there has been no formal justification for the inclusion
of a vW-like term for 2D systems at zero-temperature.
In this paper, we establish the existence of a vW-like gradient correction to the 2D KE density functional at 
{\em finite temperatures}.
Our approach parallels the earlier work of Brack,~\cite{brack} in which the ETF theory was developed in the context of ``hot'' nuclear matter (ETFT).   Given the recent work of Eschrig,~\cite{eschrig} which aims at providing a rigorous foundation
for DFT at finite-temperature, the results presented in this paper are immediately relevant to future applications of
$T>0$ DFT in low dimensional electronic systems.

The rest of our paper is organized as follows.   In the next section we provide a brief review of the general ETFT approach, followed by an
explicit calculation of the $T>0$ second order gradient correction to the 2D KE functional.  In Sec.~III we
numerically investigate the quality of the gradient corrected functional by comparing it to known, exact results,
for an isotropic 2D harmonic oscillator at finite temperature.
The paper concludes in Sec.~IV with a brief summary and suggestions for future investigations.

\section{Extended Thomas-Fermi theory at finite-temperature}
In this section, we provide a brief review of the ETFT approach.  The interested reader should refer to the 
Ref.~[\onlinecite{brack_bhaduri}] for a detailed discussion of the ETFT and the Wigner-Kirkwood semiclassical expansion.
\subsection{Semiclassical spatial density}
At the heart of the ETFT approach is the  Wigner-Kirkwood (WK) semiclassical expansion of the zero-temperature, diagonal Bloch density matrix (BDM), which in 2D is given by~\cite{brack_bhaduri}
\begin{equation}
 C_{0}(\bfr;\beta) = \left (\frac{1}{\lambda}\right)^2 e^{-\beta V(\bfr) }
 \left(
 1 - \frac{\hbar^2 \beta^2}{12 m}\left[ \nabla^2 V - \frac{\beta}{2}(\nabla V)^2\right] + \cdot \cdot \cdot
 \right)~,
\end{equation}
where $V(\bfr)$ is a local one-body potential, and for our purposes, we have only shown terms up to relative order $\hbar^2$ and $\lambda \equiv (2\pi\hbar^2 \beta/m)^{1/2}$.  Note that $\beta$ is to be viewed as a complex parameter here, and not the
inverse temperature $1/(k_BT)$.  In order to incorporate finite temperatures into the WK semiclassical theory, the finite-temperature BDM is defined by~\cite{brack}
\begin{equation}
C_{T}(\bfr;\beta) \equiv C_0 \frac{\pi \beta k_B T}{\sin (\pi \beta k_B T)}~.
\end{equation}
The finite-temperature spatial density is then obtained from an (all two-sided) inverse Laplace transform (ILT) of the finite-temperature BDM, {\it viz.,}
\begin{equation}
\rho(\bfr;T) = {\cal L}_{\mu}^{-1}\left[ 2 \frac{C_{T}(\bfr;\beta)}{\beta}\right]~,
\end{equation}
where the factor of two in Eq.~(3) accounts for the spin degeneracy after the spin trace has been taken, and $\mu$ has the physical
significance of the chemical potential.  It should be noted that if the {\em exact} $C_T(\bfr;\beta)$ is known, then Eq.~(3) will yield the exact, quantum mechanical, finite-temperature spatial density.  Of course,
here, we are using a semiclassical expansion for $C_T(\bfr;\beta)$, so the resulting $\rho(\bfr;T)$ will be the semiclassical spatial density.

In what follows, we will make use
of the following ILTs:
\begin{equation}
{\cal L}_{\mu}^{-1}\left[\beta^{n}e^{-\beta V}\frac{\pi k_B T}{\sin{(\pi \beta k_B T)}}\right] =\int_{-\infty}^{\infty} \delta^{(n)}({\tau}) \frac{1}{e^{{\tau}/k_BT}z^{-1}+1}d{\tau}~,~~~(n\ge0)
\end{equation}
\begin{equation}
{\cal L}_{\mu}^{-1}\left[\beta^{-1}e^{-\beta V}\frac{\pi k_B T}{\sin{(\pi \beta k_B T)}}\right] =\int_{0}^{\infty}  \frac{1}{e^{{\tau}/k_BT}z^{-1}+1}d{\tau}=k_B T \ln{(1+z)}
\end{equation}
\begin{equation}
{\cal L}_{\mu}^{-1}\left[\beta^{-2}e^{-\beta V}\frac{\pi k_B T}{\sin{(\pi \beta k_B T)}}\right] =\int_{0}^{\infty}  \frac{{\tau}}{e^{{\tau}/k_BT}z^{-1}+1}d{\tau}=-\left(k_BT\right)^2{\rm Li}_2(-z)
\end{equation}
where ${\rm Li}_2(\cdot)$ is the polylog function,~\cite{note3} and $z \equiv \exp[(\mu - V)/k_BT]$.   Using Eq.~(3), along with Eqs.~(4--6), readily leads to the following second-order expression for the finite-temperature spatial density:
\begin{eqnarray}
\rho(\bfr;T) &=& \left(\frac{mk_B T}{\pi\hbar^2}\right) \ln{(1+z)} -\frac{\left(\nabla V\right)^2}{24\pi k_B^2T^2}\frac{z(z-1)}{(z+1)^3}- \frac{\nabla^2 V}{12\pi k_BT}\frac{z}{(z+1)^2}\nonumber \\
&=& \rho^{(0)}(\bfr;T) + \rho^{(2)}(\bfr;T)~.
\end{eqnarray}
The ETFT density in Eq.~(7) is well-defined throughout all space, with the last two terms, denoted by $ \rho^{(2)}(\bfr;T)$, being relative order $\hbar^2$ greater
than the first term, $ \rho^{(0)}(\bfr;T)$.
Note that in Eq.~(7), for $V(\bfr)< \mu$, $z\to \infty$ {\em exponentially} as $T\to 0$, so that the $T\to 0$ limit is well-defined 
only {\em within the classical region}; 
the non-analytic behaviour of the zero-temperature ETF densities at the turning point, $\mu=V(\bfr)$, is well-known.~\cite{brack_bhaduri}  What we 
find here is that the singular behaviour of the $T=0$ densities 
cannot be avoided by first formulating the semiclassical theory at finite-temperature, 
and then performing
the $T\to 0$ limit.~\cite{brack2}  We have, however, confirmed that the $T\to 0$ limit of Eq.~(7) with $V(\bfr)<\mu$
correctly reduces to the 
known (albeit problematic) $T=0$ result.~\cite{brack_bhaduri}
\subsection{Semiclassical kinetic energy density}
The KE density may be obtained from knowledge of the finite-temperature first-order density matrix (FDM).  To this end, it is useful to 
introduce the centre-of-mass, 
 $\bfq=(\bfr + \bfr')/2$, and relative coordinates, $\bfs=\bfr - \bfr'$, so that we may write three variants of the KE density:~\cite{shea_vanzyl,time_reversal_symmetry}
\begin{equation}
{\cal T}(\bfr;T) =- \frac{\hbar^2}{2m} \left( \frac{1}{4} \nabla_{\bfq}^2 + \nabla_{\bfs}^2\right)\rho(\bfq,\bfs;T)|_{\bfs = 0}~,
\end{equation}
\begin{equation}
{\cal T}_1(\bfr;T) =\frac{\hbar^2}{2m} \left( \frac{1}{4} \nabla_{\bfq}^2 - \nabla_{\bfs}^2\right)\rho(\bfq,\bfs;T)|_{\bfs = 0}~,
\end{equation} 
\begin{equation}
\xi(\bfr;T) = \frac{{\cal T}(\bfr;T) + {\cal T}_1(\bfr;T)}{2} = -\frac{\hbar^2}{2m}\nabla^2_{\bfs}\rho(\bfq,\bfs;T)|_{\bfs = 0}~.
\end{equation}
Again, if the exact finite-temperature expression for the FDM is known, then Eqs.~(8--10) will yield the exact, quantum mechanical finite-temperature KE density.

While all three of the above expressions for the KE density integrate to the exact same kinetic energy, ${\cal T}_1(\bfr)$ is strictly positive definite, and is therefore sometimes preferred in applications of 
density functional theory.  It has already been shown long ago that 
${\cal T}(\bfr;T)$ and ${\cal T}_1(\bfr;T)$ generally have oscillations exactly opposite in phase, so that their mean, $\xi(\bfr;T)$, is a smooth function.  In this paper, we focus on ${\cal T}(\bfr;T)$ in order
to make contact with earlier theoretical work done at zero-temperature, where the {\em exact} ${\cal T}(\bfr;T=0)$ was investigated (see also Eq.~(26) in Sec.~III).~\cite{brack_vanzyl}

The kinetic energy density in Eq.~(8) may also be expressed in terms of only local quantities, {\it viz.,}~\cite{brack}
\begin{equation}
{\cal T}(\bfr;T) = -\rho(\bfr;T)V(\bfr)+\mathfrak{F}(\bfr;T)+T\sigma(\bfr;T)~, 
\end{equation}
where
\begin{equation}
\mathfrak{F}(\bfr;T)= \mu\rho(\bfr;T) - {\cal L}_{\mu}^{-1}\left[\frac{C_T}{\beta^2}\right]~,
\end{equation}
is the free energy density, and
\begin{equation}
\sigma(\bfr;T) = \frac{\partial}{\partial T} {\cal L}_{\mu}^{-1}\left[\frac{C_T }{\beta^2}\right]~,
\end{equation}
is the entropy density.  The semiclassical approximation to ${\cal T}(\bfr;T)$ may easily be determined to second-order by employing the semiclassical approximation
to $C_T(\bfr;\beta)$, as in Sec. IIA for the spatial density.
A straightforward calculation results in the following expression for the finite-temperature, semiclassical KE density
\begin{eqnarray}
{\cal T}(\bfr;T) &=& -\frac{m k_B^2T^2}{\pi \hbar^2}{\rm Li}_2(-z)- \frac{z\left(\nabla V\right)^2}{12\pi k_BT(z+1)^2}+\frac{\nabla^2 V}{12\pi}\frac{z}{z+1} \nonumber \\
&\equiv& {\cal T}^{(0)}(\bfr;T) + {\cal T}^{(2)}(\bfr;T)~,
\end{eqnarray}
where, following Eq.~(7), the last two terms in Eq.~(14) are denoted collectively by ${\cal T}^{(2)}(\bfr;T)$.  As in 
Eq.~(7), the $T\to 0$ limit of Eq.~(14) is well-defined 
only within the classical region.  
Finally, it follows immediately from Eqs.~(8) and (9) that
\begin{equation}
 {\cal T}_1(\bfr;T) = {\cal T}(\bfr;T)+ \frac{\hbar^2}{4m}\nabla^2\rho~.
 \end{equation}
\subsection{Second-order kinetic energy density functional}
For the special case of 2D, the elimination of $z$ and $V$ in  ${\cal T}(\bfr;T)$ above, in favour of the spatial density, $\rho$, is quite
straightforward.\cite{note1,brack2,ring}  
To begin we define ${\tilde{V}}\equiv \frac{\mu-V}{k_BT}$, so that $z = \exp({\tilde V})$. 
Thus, the density, Eq.~(7), is a function of ${\tilde V}$, $\nabla {\tilde V}$ and $\nabla^2{\tilde V}$.
Calculating from Eq.~(7) $\nabla \rho$ and $\nabla^2\rho$ and consistently neglecting higher than second derivatives
of the potential, we have: $\rho=\rho(\tV,\nabla\tV,\nabla^2\tV)$, $\nabla\rho=\nabla\rho(\cdot\cdot\cdot)$, and
$\nabla^2\rho=\nabla^2\rho(\cdot\cdot\cdot)$, which can be solved for $\tV, \nabla\tV$ and $\nabla^2\tV$; inserting this
into Eq.~(14) yields the finite-temperature kinetic energy density functional up to ${\cal O}(\hbar^2)$, {\it viz.,}
 \begin{eqnarray}
 {\cal T}_{\rm ETFT}[\rho] &=& -A_Tk_BT{\rm Li}_2(1-e^{\rho/A_T})-
 \frac{\hbar^2}{12m}\nabla^2 \rho-
 \frac{\hbar^2}{12m}f_1(\bfr;T) \nabla^2\rho\nonumber \\
 &+& \frac{\hbar^2}{24m}f_2(\bfr;T) \frac{(\nabla \rho)^2}{\rho}\nonumber \\
 &=& {\cal T}_{\rm TFT}[\rho] + {\cal T}^{(2)}_{\rm ETFT}[\rho]~,
 \end{eqnarray}
 where 
 \begin{equation}
f_1(\bfr;T) = \frac{\rho}{A_T(e^{\rho/A_T}-1)}~,
 \end{equation}
 \begin{eqnarray}
f_2(\bfr;T) = e^{\rho/A_T}[f_1(\bfr;T)]^2~,
 \end{eqnarray}
 and $A_T \equiv m k_BT/(\pi\hbar^2)$.
 The first term in Eq.~(16) is the finite-temperature 2D TFT KE density functional, ${\cal T}_{\rm TFT}[\rho]$, while the other three terms
represent the ${\cal O}(\hbar^2)$ gradient corrections.  As advertised, the last term in 
 Eq.~(16) has the vW form $\sim (\nabla\rho)^2/\rho$.    
 
An explicit expression for the ${\cal T}_1[\rho]$ KE functional may also be given by making use of Eq.~(15), {\it viz.,}
 \begin{eqnarray}
 {\cal T}_{1,{\rm ETFT}}[\rho] &=& -A_Tk_BT{\rm Li}_2(1-e^{\rho/A_T})+\frac{\hbar^2}{6m}\nabla^2 \rho-\frac{\hbar^2}{12m} f_1(\bfr;T)\nabla^2\rho\nonumber \\
 &+& \frac{\hbar^2}{24m}f_2(\bfr;T) \frac{(\nabla \rho)^2}{\rho}~.
 \end{eqnarray}
Recall that ${\cal T}_{\rm ETFT}[\rho]$ and ${\cal T}_{\rm 1,ETFT}[\rho]$ both integrate to the same total kinetic energy for finite systems since the Laplacian
 term is the divergence of a vector field which vanishes at infinity, and by Gauss's theorem will not contribute the kinetic energy.  
 
To investigate the low temperature behaviour of Eqs.~(16) and (19) we use,
 \begin{equation}
 \lim_{\rho/A_T \to \infty}{\rm  Li}_2\left(-e^{\rho/A_T}\right) = -\frac{1}{2}\left(\frac{\rho}{A_T}\right)^2~,
 \end{equation}
 along with the fact that $f_1(\bfr;T)\to 0$, and $f_2(\bfr;T)\to 0$ as $T \to 0$, to write 
  \begin{equation}
 {\cal T}_{\rm ETFT}[\rho] \to \frac{\hbar^2}{2m}\left(\pi \rho^2-\frac{1}{6}\nabla^2 \rho\right)~,
 \end{equation}
 and
 \begin{equation}
 {\cal T}_{1,{\rm ETFT}}[\rho] \to \frac{\hbar^2}{2m}\left(\pi \rho^2+\frac{1}{3}\nabla^2 \rho\right)~.
 \end{equation}
Equations (21) and (22) agree with the known results for the $T=0$ 2D KE functionals ${\cal T}[\rho]$ and ${\cal T}_1[\rho]$, respectively.~\cite{brack_bhaduri}
As mentioned above, for physical densities, integration over $\nabla^2 \rho$ vanishes.  Therefore, in any practical
implementation of self-consistent $T=0$ DFT, the Laplacian term may be ignored, and we may write
 \begin{equation}
 {\cal T}_{\rm ETFT}[\rho] = {\cal T}_{1,{\rm ETFT}}[\rho] =  \frac{\hbar^2}{2m}\left(\pi \rho^2\right)~,
 \end{equation}
 as $T\to0$.
 We see that there is no vW-like gradient correction at $T=0$, leading to the incorrect conclusion that for an inhomogeneous 2D Fermi gas, the TF KE functional (at least to second-order) is exact.~\cite{note2}
 We would like to stress again that the non-uniqueness of the KE density does not alter the result that there are no vW-like gradient corrections at $T=0$; the only differences between the $T=0$ semiclassical KE densities obtained from Eqs.~(8--10) are Laplacian terms which, as we have already stated, are
of no consequence since they vanish upon integration for physical (i.e., finite) systems. 

It is readily found that as $T\to \infty$, $f_1(\bfr;T) \to 1$ and $f_2(\bfr;T) \to 1$, so that up to ${\cal O}(\hbar^2)$, the
2D KE functionals go over to
\begin{equation}
{\cal T}_{\rm ETFT}[\rho] \to {\cal T}_{B}[\rho] = k_BT\rho - \frac{\hbar^2}{6m}\nabla^2\rho + 
\frac{\hbar^2}{24m}\frac{(\nabla \rho)^2}{\rho}~,
\end{equation}
and
\begin{equation}
{\cal T}_{1,{\rm ETFT}}[\rho] \to {\cal T}_{1,B}[\rho]= k_BT\rho + \frac{\hbar^2}{12m}\nabla^2\rho + 
\frac{\hbar^2}{24m}\frac{(\nabla \rho)^2}{\rho}~.
\end{equation}
Therefore, at high-temperature, the second order gradient corrections take a functional form analogous to what is found in 3D ETF,
although the numerical pre-factors are different.  It is also interesting to note that the high-temperature limit of the TFT term is linear
in the density,
$k_BT\rho$, in contrast to the quadratic dependence, $\pi\rho^2/2$, exhibited in the zero-temperature limit.  We have also checked that the high-temperature limit of the 2D KE functional may be obtained by  performing 
an analogous calculation assuming a Boltzmann, rather than a Fermi, gas.  
\section{Comparison with exact results}
In a previous study, Brack and van Zyl~\cite{brack_vanzyl} examined the $T=0$ 2D TF KE functional by comparing its global
({\it i.e.,} integrated) and local ({\it i.e.,} spatially dependent)  properties with the known analytical expressions for the  2D harmonic
oscillator (HO) potential.  They found the remarkable result that at $T=0$, the  2D TF KE functional ({\em without gradient corrections}), when using
the exact spatial density of the 2D HO, gives the exact quantum mechanical kinetic energy.  This result is highly non-trivial
because the TF functional is simply the LDA to the true KE, and therefore, cannot be variationally exact.    More
surprising, however, is how well the local behaviour of the exact KE density is reproduced by the TF approximation, as
illustrated in Fig.~3 of Ref.~[\onlinecite{brack_vanzyl}].
The purpose of this section is to perform an analogous calculation for the finite-temperature KE density functionals
presented in this paper.  In our numerical calculations, we have scaled all energies and lengths
by  $\hbar\omega$, and $\ell_{\rm osc} = \sqrt{\hbar/m\omega}$, respectively.
We have also restricted our attention to relatively small particle numbers, $N$, since it can be shown rigorously that in the
large-$N$ limit, the TF approximation becomes exact.~\cite{vanzyl_suzuki}

The exact finite-temperature kinetic energy density is given by (see also Eq.~(8))
\begin{equation}
{\cal T}_{\rm exact}(\bfr;T) = -\frac{1}{2}\left(\frac{1}{4}\nabla^2_{\bfq} + \nabla^2_{\bf s}\right) \rho_{\rm exact}(\bfq,\bfs;T)|_{\bfs = 0}~,
\end{equation}
where specializing to the case of the  2D HO, 
\begin{equation}
\rho_{\rm exact}(\bfq,\bfs;T) = \frac{2}{\pi}\sum_{n=0}^{\infty}\sum_{k=0}^{\infty} (-1)^n L_n(2q^2)L_k(s^2/2)e^{-(q^2+s^2/4)}
\frac{1}{\exp(\frac{n+1+k-\mu}{T})+1}~,
\end{equation}
is the {\em exact} finite-temperature first-order density matrix.  Putting $\bfs = 0$ in Eq.~(27) gives the exact finite-temperature particle density
for the 2D HO, {\it viz.,}
\begin{equation}
\rho_{\rm exact}(\bfr;T) = \frac{2}{\pi}\sum_{n=0}^{\infty}\sum_{k=0}^{\infty} (-1)^n L_n(2r^2)e^{-r^2}
\frac{1}{\exp(\frac{n+1+k-\mu}{T})+1}~.
\end{equation}

In Table I below, we present a numerical comparison of the kinetic energies as obtained from
\begin{eqnarray}
K_{\rm exact} &=& \int{\cal T}_{\rm exact}(\bfr) d^2r~,~~~~~~~~{\rm c.f. ~Eq.~(23)} \\
K_{\rm TFT} &=& \int{\cal T}_{\rm TFT}[\rho_{\rm exact}] d^2r,~~~~{\rm c.f.~ Eq.~(12)} \\
K_{\rm ETFT} &=& \int{\cal T}_{\rm ETFT}[\rho_{\rm exact}] d^2r,~~~{\rm c.f.~ Eq.~(12)}~,
\end{eqnarray}
for $N=42$ particles, with Table II providing the same calculation for $N=420$ particles.
\begin{table}[h]
\begin{center}
\begin{tabular}
{|p{0.8cm}|p{2.cm}|p{2.cm}|p{2.cm}|p{2.cm}||}
\hline
$T$ & $K_{\rm exact}$ &$K_{\rm TFT}$ &$K_{\rm ETFT}$\\ \hline\hline
$0.5$ & 93.8984	&     93.4202	&   93.6977\\ \hline
$0.8$  & 97.8489	&     97.3152	&   97.7439  \\ \hline
$1.0$ &  101.3291	&    100.7624   &   101.2613  \\ \hline
$2.0$  &125.9995	&    125.3372	&  125.9854  \\ \hline
$3.0$ & 157.9066	&    157.2537	&  157.9007 \\ \hline
$4.0$ & 193.5600	&    192.9616	&  193.5591   \\ \hline
$5.0$ & 231.2762	 &   230.6358	&  231.1417 \\ \hline
\end{tabular}
\end{center}
\caption{A comparison of the total kinetic energy at various temperatures as determined from Eqs.~(29--31) with $N=42$ particles.  All quantities
are measured in scaled units, as discussed in the text.  The largest relative percentage errors in our tabulated data are
$\Delta K_{\rm TFT}\sim 0.5\%$ and $\Delta K_{\rm ETFT}\sim 0.2\%$.}
\end{table}
\begin{table}[h]
\begin{center}
\begin{tabular}
{|p{0.8cm}|p{2.cm}|p{2.cm}|p{2.cm}|p{2.cm}||}
\hline
$T$ & $K_{\rm exact}$ &$K_{\rm TFT}$ &$K_{\rm ETFT}$ \\ \hline\hline
$0.5$ & 2879.2573	  &      2877.8822    &	  2878.3566    \\ \hline
$0.8$ &  2892.3488	   &     2890.9642    &	  2891.7668 \\ \hline
$1.0$ & 2904.3701	  &      2902.9579	&  2903.9484 \\ \hline
$2.0$ &3002.5299	  &      3000.9349	&  3002.4338 \\ \hline
$3.0$ & 3158.2817	 &       3156.5131   &	  3158.2465 \\ \hline
$4.0$ &3361.6954	  &      3359.7806	&  3361.6754 \\ \hline
$5.0$ &3603.0235	 &       3601.0015	&  3603.0076  \\ \hline
\end{tabular}
\end{center}
\caption{As in Table I, but with $N=420$ particles. The largest relative percentage errors in our tabulated data are
$\Delta K_{\rm TFT}\sim 0.06\%$ and $\Delta K_{\rm ETFT}\sim 0.03\%$.}
\end{table}
\begin{figure}[]
   {\includegraphics[angle=0, width=136mm]{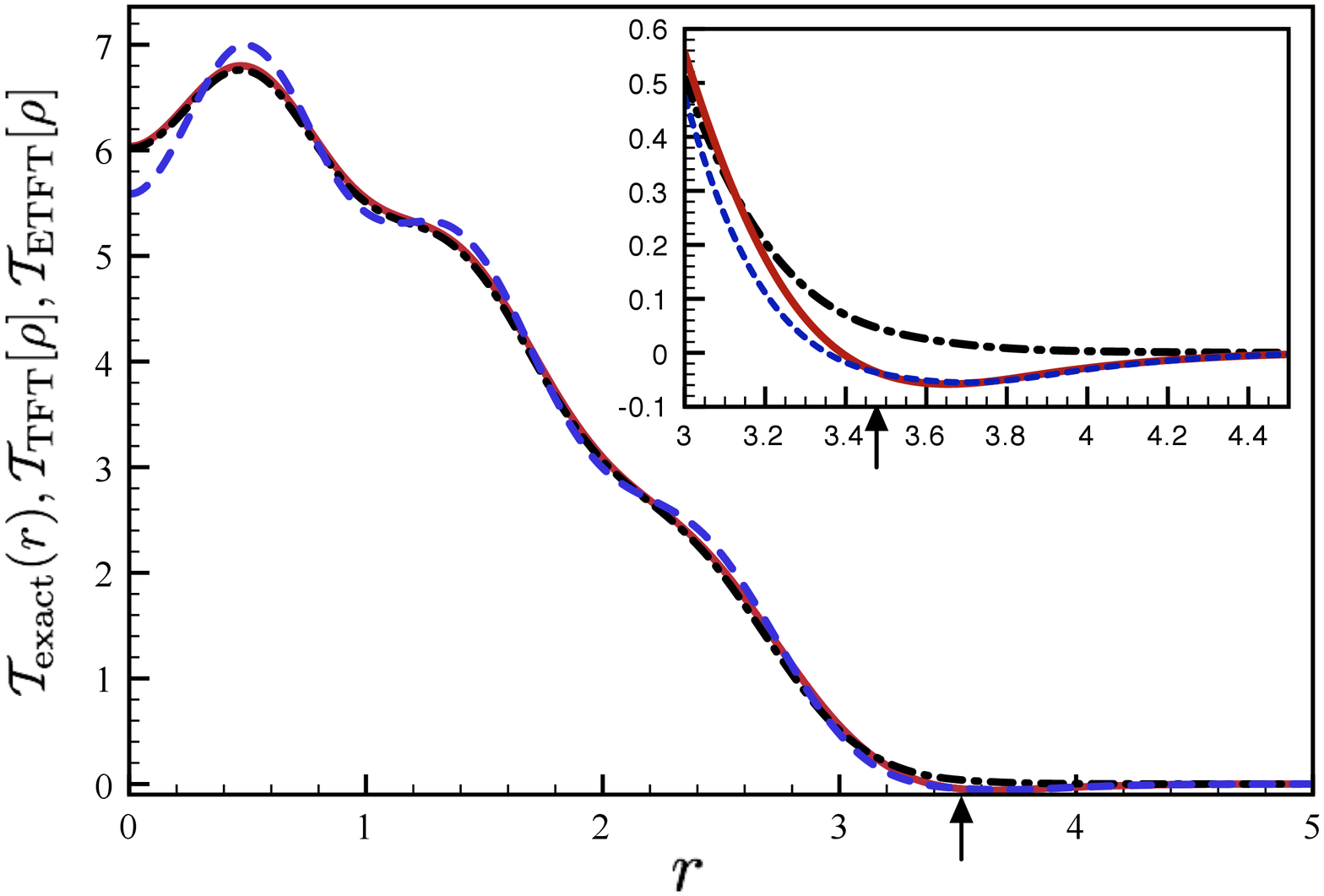}}
   \caption{\label{fig2}%
            (colour online)  A plot of the exact, Eq.~(26), TFT, {\it i.e.,} the first term in Eq.~(16), and the ETFT, Eq.~(16), 
            KE densities for $N=42$ particles and $T=0.2$.  The solid (red online) curve is the exact KE density, the dot-dashed (black online) curve is the TFT KE density,
            and the dashed (blue online) curve is the ETFT KE density.
            Inset: Magnification of the tail region,
            where deviations between the three curves are most pronounced.  The arrow indicates the classical turning point. Scaled units are used, as discussed in the text.}
  \end{figure}
 
It is clear that the TFT KE, $K_{\rm TFT}$, is always lower than the exact KE density, $K_{\rm exact}$, while the gradient
corrections serve to improve the agreement with the exact result. 
This makes sense given that the gradient
corrections take into account the curvature of the system imposed by the external potential, thereby increasing
the kinetic energy of the system.  It is nevertheless quite surprising how well the TFT functional does in 
describing the kinetic energy of the strongly inhomogeneous 2D HO at finite temperature, even for small particle numbers. 
From our numerical calculations, we observe that a ten-fold increase in the number of particles reduces the {\em largest} relative percentage error by roughly a factor of ten; the better agreement between the TFT, ETFT and the exact KE, is in keeping with the expected result that in the large-$N$ limit, the TF approximation becomes exact.

\begin{figure}[]
   {\includegraphics[angle=0, width=136mm]{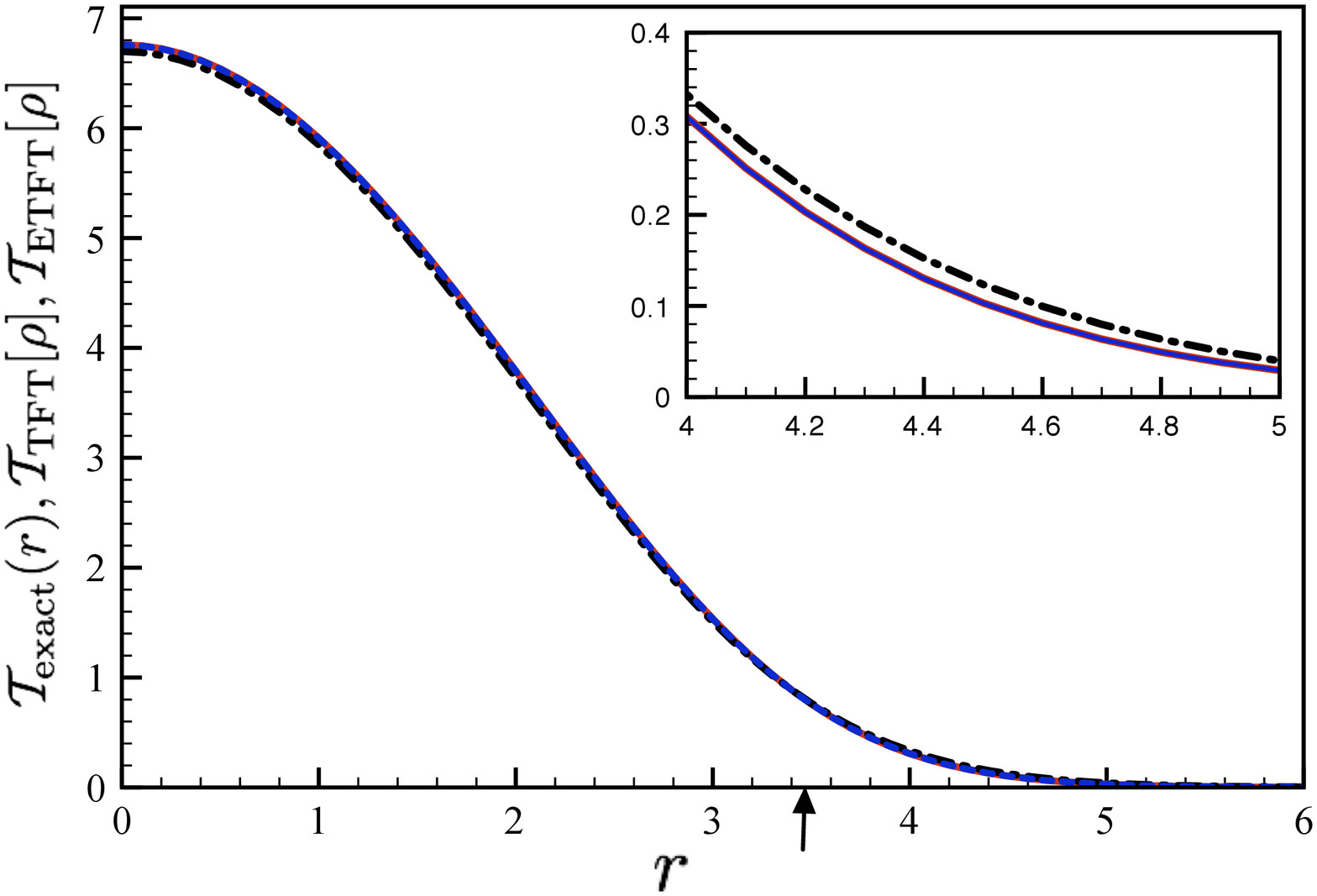}}
   \caption{\label{fig3}%
            (colour online)  As in Fig.~\ref{fig2}, but with $T=2.0$.   Note that at this temperature,  the shell oscillations are
            already completely washed away.  Inset:  Magnification of the tail region, clearly illustrating that the exact (solid curve, red online) and ETFT (dashed curve, blue online)
            curves are indistinguishable on the scale of the plot. The arrow indicates the classical turning point.}
  \end{figure}

In Figs.~\ref{fig2} and \ref{fig3}, we present the KE densities, ${\cal T}_{\rm exact}(\bfr)$,  ${\cal T}_{\rm TFT}[\rho]$, and 
${\cal T}_{\rm ETFT}[\rho]$ with $N=42$ particles, at $T=0.2$ and $T=2$, respectively. As in Tables I and II, the exact spatial density, Eq.~(28), has been used as
input for the KE functionals.  
 We have focused on a small number of particles, {\it viz.,} $N=42$, since deviations between the exact, TFT, and ETFT densities
 are more pronounced for $N \lesssim {\cal O}(10^2)$, particularly at low temperatures. 
 
Figure \ref{fig2} reveals several interesting aspects of the level of approximation at low temperatures.  First, we note that the TFT
(dot-dashed curve, black online) and the exact KE density (solid curve, red online) are almost indistinguishable within $0 < r \lesssim 3$.  However, near the
tail region (see figure inset), it is clear that the two KE densities are quite different; the TFT density is strictly positive definite, whereas
the exact KE density falls below zero, before coalescing with the TFT density for $r\gtrsim 4.5$.  
The ETFT KE density (dashed curve, blue online), on the other hand, does a relatively poor job of quantitatively capturing the behaviour of the exact KE
density in the interior, but for the tail region, more accurately
reproduces the exact result.  Indeed, for $r\gtrsim 3.5$, the ETFT and exact KE densities are indistinguishable
on the scale of the inset.  
Moreover, in spite of the differences between the exact and ETFT densities for $r \lesssim 3$, the ETFT
is still a better approximation for the total KE, as evidenced by the data presented in Table I.  Therefore, while the gradient corrections
are important for improving the total ({\it i.e.,} integrated) KE, they are essential for describing the correct low-temperature behaviour of the exact KE past the classical turning point.

Figure \ref{fig3} presents the same data as in Fig.~\ref{fig2}, but with $T=2.0$.  At this temperature, the shell oscillations are already completely washed out, and the exact and ETFT KE densities are {\em indistinguishable} from each other, including the tail region; in the inset, there are actually three curves plotted, but the difference between the solid (red online) and dashed (blue online) curves cannot be
resolved. Thus, at temperatures for which the shell effects are absent ({\it i.e.,} $T \gtrsim 1$), the ETFT 
is an excellent  approximation to the exact finite-temperature KE density.  Near the tail region, we see that the TFT KE density (dot-dashed curve, black online) is consistently too
large, thereby emphasizing the importance of the  gradient corrections for a faithful description of the local behaviour of the exact KE density, even for small particle numbers.

\begin{figure}[]
   {\includegraphics[angle=0, width=146mm]{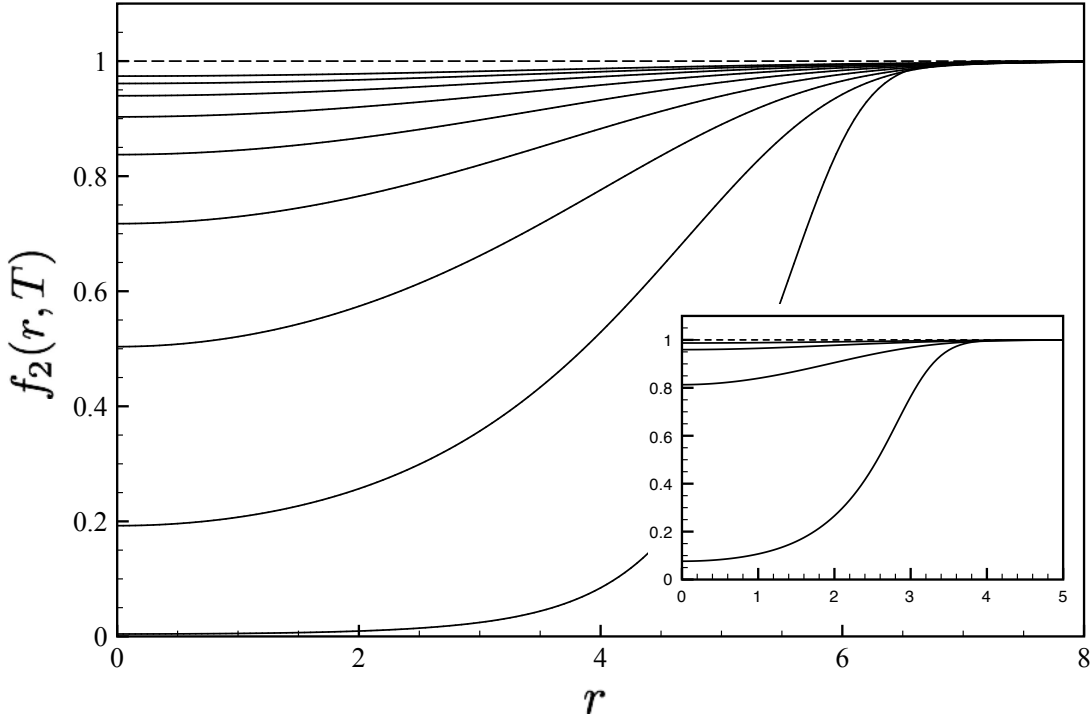}}
   \caption{\label{fig1}%
            A plot of the coefficient of the von Weizs\"acker term, Eq.~(18), for $N=420$ particles.  The curves correspond, from
            lowest to highest, to $T=2$ and $T=18$, respectively, in steps of $\Delta T=2$.  Note that by $T=18$, $f_2(\bfr;T)$ is already approaching its  Boltzmann value, $f_2(\bfr;T)=1$, represented by the dashed line. 
            Inset: As in the main figure, but for  $N=42$ particles at temperatures $T=1, 3, 5, 7$.  Scaled units have been used, as discussed
            in the text.}
  \end{figure}

Finally, in Fig.~\ref{fig1}, we illustrate the temperature and
spatial dependence of the vW coefficient, $f_2(\bfr;T)$ given by Eq.~(18), for $N=420$ particles, and $N=42$ (inset).  As the temperature is increased, we see that the vW coefficient approaches
the constant value $f_2(\bfr;T)=1$, confirming our analytical results above for the Boltzmann regime.  Figure \ref{fig1} establishes that $T \gtrsim 18$ is a sufficiently
high temperature for the $N=420$ particle system to be treated as Boltzmann gas.  The inset illustrates the expected result that for smaller particle numbers, one enters the Boltzmann regime at much lower temperatures.
\section{Conclusions and Outlook}
We have provided a formal justification for the inclusion of gradient corrections to the 2D KE density functional of an ideal Fermi gas at finite temperatures.  
Our numerical calculations have  examined the quality of the TFT and ETFT functionals by comparing them against exact, analytical results for
the 2D HO potential.  We find that  gradient corrections lead to an improved agreement for total KE when compared to the TFT approximation, and are necessary  
to correctly reproduce the quantum mechanical tunneling into the classically forbidden region exhibited by the exact KE density.
Unfortunately, the non-analytic behaviour of the $T=0$ semiclassical densities at the classical turning point cannot be remedied 
within the present formalism.  

An extension of this work would be to develop the finite-temperature Dirac exchange functional, which could then be used in a fully self-consistent,
finite-temperature Thomas-Fermi-Dirac von Weizs\"acker (TFDW) DFT calculation similar to what has already been done
at $T=0$ for low-dimensional Fermi systems.~\cite{vz1}  It would be interesting to see whether the optimal, {\it ad hoc}, $T=0$ vW coefficient of $1/8$ could be motivated from a finite-temperature self-consistent 
TFDW calculation.

Finally, we wish to point out that the results presented here may also find relevance in current experiments on ultra-cold, trapped Fermi gases, in which
inter-atomic interactions may be tuned from essentially zero, to very strong, {\it via} the
Feshbach resonance.~\cite{hulet}  It is possible that the low-temperature shell oscillations, and their suppression as the temperature is increased, 
may be directly observable in cold atoms experiments on low-dimensional systems.

 \acknowledgments
 B. P. van Zyl would like to acknowledge financial support from the Natural Sciences and Engineering Research Council (NSERC) of Canada through the Discovery Grants program.
 A. Farrell would like to acknowledge the NSERC Undergraduate Student Research Awards (USRA) program for additional 
 financial support.



\begin{thebibliography}{99}
\bibitem{note3}
The polylog function, ${\rm Li}_2(\cdot)$ is identical to the Fermi integral, $F_{\xi} = \int_0^{\infty}\frac{\tau^{\xi}}{1+\exp(\tau-\eta)} d\tau$,
with $\xi = 1$.  See Ref.~[\onlinecite{brack_bhaduri}] for details.
\bibitem{note1}  The functional inversion procedure used here is not possible in 3D because of the complicated dependence of $z$ and $V$
on the spatial density.   We refer the reader to Ref.~[\onlinecite{brack2}] for a detailed derivation of the elimination of $z$ and $V$ in favour
of $\rho$ to obtain the 3D finite-temperature KE functional.
\bibitem{note2}
While we have not presented the cumbersome calculation here, the more general result that {\em all} finite-temperature gradient corrections vanish at $T=0$ has also been established.
\end{thebibliography}
\end{document}